\documentclass[superscriptaddress,twocolumn,10pt,prx,aps,showpacs,longbibliography,floatfix]{revtex4-2}

\usepackage[dvipsnames]{xcolor}

\definecolor{darkred}{rgb}{0.6,0.05,0.05}
\definecolor{darkgreen}{rgb}{0.05,0.6,0.05}
\definecolor{darkblue}{rgb}{0.05,0.05,0.6}
\usepackage[colorlinks]{hyperref}
\hypersetup{colorlinks,
linkcolor=darkred,
filecolor=darkgreen,
urlcolor=darkblue,
citecolor=darkgreen}

\usepackage{textcomp,mathcomp}
\usepackage{mathtools}
\usepackage{amsmath}
\usepackage{amssymb}
\usepackage[normalem]{ulem}
\usepackage{amsthm}
\usepackage{xfrac}
\usepackage{dsfont}
\usepackage{siunitx}
\usepackage{physics}
\usepackage{graphicx}
\usepackage{hyperref}
\usepackage{dcolumn}
\usepackage{bm}

\usepackage{multirow}
\usepackage{tabularx}
\usepackage{booktabs}

\usepackage{soul}

\definecolor{armygreen}{rgb}{0.29, 0.33, 0.13}
\definecolor{palatinatepurple}{rgb}{0.41, 0.16, 0.38}
\definecolor{sangria}{rgb}{0.57, 0.0, 0.04}

\newcommand{\rhot}{\rho}
\newcommand{\sss}{\rho_{\rm ss}}

\makeatletter
\newcommand*\bigcdot{\mathpalette\bigcdot@{.5}}
\newcommand*\bigcdot@[2]{\mathbin{\vcenter{\hbox{\scalebox{#2}{$\m@th#1\bullet$}}}}}
\makeatother

\renewcommand{\paragraph}[1]{\vspace{5pt}\noindent \textbf{#1--.}}

\newcommand{\detuning}{\delta}
\newcommand{\discretization}{\epsilon}
\newcommand{\average}[1]{\expval{#1}}

\newcommand{\standarddeviation}[1]{\Delta #1}
\newcommand{\error}[1]{\Delta  #1}

\begin{document}

\author{Guillaume Beaulieu}
\affiliation{Hybrid Quantum Circuits Laboratory (HQC), Institute of Physics, \'{E}cole Polytechnique F\'{e}d\'{e}rale de Lausanne (EPFL), 1015 Lausanne, Switzerland}
\affiliation{Center for Quantum Science and Engineering, \\ \'{E}cole Polytechnique F\'{e}d\'{e}rale de Lausanne (EPFL), CH-1015 Lausanne, Switzerland}
\author{Fabrizio Minganti}
\thanks{Presently at Alice \& Bob, Paris, France}
\affiliation{Center for Quantum Science and Engineering, \\ \'{E}cole Polytechnique F\'{e}d\'{e}rale de Lausanne (EPFL), CH-1015 Lausanne, Switzerland}
\affiliation{Laboratory of Theoretical Physics of Nanosystems (LTPN), Institute of Physics, \'{E}cole Polytechnique F\'{e}d\'{e}rale de Lausanne (EPFL), 1015 Lausanne, Switzerland}
\author{Simone Frasca}
\affiliation{Hybrid Quantum Circuits Laboratory (HQC), Institute of Physics, \'{E}cole Polytechnique F\'{e}d\'{e}rale de Lausanne (EPFL), 1015 Lausanne, Switzerland}
\affiliation{Center for Quantum Science and Engineering, \\ \'{E}cole Polytechnique F\'{e}d\'{e}rale de Lausanne (EPFL), CH-1015 Lausanne, Switzerland}
\author{Marco Scigliuzzo}
\affiliation{Center for Quantum Science and Engineering, \\ \'{E}cole Polytechnique F\'{e}d\'{e}rale de Lausanne (EPFL), CH-1015 Lausanne, Switzerland}
\affiliation{Laboratory of Photonics and Quantum Measurements (LPQM),  Institute of Physics, EPFL, CH-1015 Lausanne, Switzerland}
\author{Simone Felicetti}
\affiliation{Institute for Complex Systems, National Research Council (ISC-CNR),Via dei Taurini 19, 00185 Rome, Italy}
\affiliation{Physics Department, Sapienza University, P.le A. Moro 2, 00185 Rome, Italy}
\author{Roberto Di Candia}
\affiliation{Department of Information and Communications Engineering, Aalto University, Espoo 02150, Finland}
\affiliation{Dipartimento di Fisica, Universit\`a degli Studi di Pavia, Via Agostino Bassi 6, I-27100, Pavia, Italy}
\author{Pasquale Scarlino}
\email[E-mail: ]{pasquale.scarlino@epfl.ch}
\affiliation{Hybrid Quantum Circuits Laboratory (HQC), Institute of Physics, \'{E}cole Polytechnique F\'{e}d\'{e}rale de Lausanne (EPFL), 1015 Lausanne, Switzerland}
\affiliation{Center for Quantum Science and Engineering, \\ \'{E}cole Polytechnique F\'{e}d\'{e}rale de Lausanne (EPFL), CH-1015 Lausanne, Switzerland}

\title{Criticality-Enhanced Quantum Sensing with a Parametric Superconducting Resonator}

\date{\today}

\begin{abstract}
Quantum metrology, a cornerstone of quantum technologies, exploits entanglement and superposition to achieve higher precision than classical protocols in parameter estimation tasks. When combined with critical phenomena such as phase transitions, the divergence of quantum fluctuations is predicted to enhance the performance of quantum sensors. Here, we implement a critical quantum sensor using a superconducting parametric (i.e., two-photon driven) Kerr resonator. The sensor, a linear resonator terminated by a superconducting quantum interference device, operates near the critical point of a finite-component second-order dissipative phase transition obtained by scaling the system parameters. We analyze the performance of a frequency-estimation protocol and show that quadratic precision scaling with respect to the system size can be achieved with finite values of the Kerr nonlinearity. Since each photon emitted from the cavity carries more information about the parameter to be estimated compared to its classical counterpart, our protocol opens perspectives for faster or more precise metrological protocols.
Our results demonstrate that quantum advantage in a sensing protocol can be achieved by exploiting a {\it finite-component} phase transition.
\end{abstract}

\maketitle

\section{Introduction}
 \label{sec:introduction}

Quantum technologies can offer significant advantages in metrology and sensing~\cite{Degen2016}. Critical quantum sensing is a promising approach that exploits the quantum properties of phase transitions to achieve optimal precision scaling~\cite{Zanardi2008,invernizzi2008Optimal,Tsang2013,ivanov_adiabatic_2013,Bina2016,macieszczak_dynamical_2016, Frerot2018}.
Theoretical studies primarily focused on quantum phase transitions~\cite{Mirkhalaf2020,Montenegro2021,Niezgoda2021,DiFresco2022,Sahoo24,montenegro2024}, which occur in closed quantum system at zero temperature when the ground state changes as a function of one of the parameters.
The non-commutative nature of Hamiltonian terms triggers the transition. 
While the transition occurs only in the thermodynamic limit of, e.g., an infinite number of atoms, the expected enhanced sensing properties already manifest when scaling-up the system towards this limit \cite{di2023critical}.

Despite critical slowing down, critical quantum sensing protocols can saturate the fundamental precision bounds~\cite{Rams2018}.
Experimentally, quantum critical sensors have been implemented using many-body Rydberg atoms~\cite{ding2022enhanced} and nuclear magnetic resonance techniques~\cite{Liu2021}.

Beyond their application in quantum computing, communication, and simulation \cite{Blais2021}, superconducting circuits have a long-standing history in sensing~\cite{Clarke1980, Jaklevic1964,quantum_sensing_with_superconducting_circuits}. Superconducting quantum interference devices (SQUIDs) are used as highly-sensitive magnetic field sensors; qubits are employed to determine the amplitude of microwave signals at cryogenic temperature through spectroscopy measurements~\cite{HniglDecrinis2020,Schuster2007}; complex junction-based architectures have been proposed as magnetic-flux sensors to probe magnetic structures~\cite{Wolski2020,LachanceQuirion2020}, detect axions, and conduct dark matter research~\cite{Dixit2021}. 
Despite theoretical proposals and some experimental evidence of quantum phase transitions in superconducting devices~\cite{Zhang2023,Houck2012}, quantum-enhanced critical sensing protocols have yet to be implemented in superconducting circuits. 
A main challenge, both in superconducting and other experimental architectures, remains the presence of classical noise and decoherence~\cite{Maccone_entangl,Gorecki_bounds}, making it challenging to scale-up these systems while maintaining the low dissipation rates necessary for observing quantum phase transitions.

Recent theoretical proposals show that critical sensing protocols can also be based on finite-component quantum phase transitions~\cite{Ashhab2013,hwang_quantum_2015,Puebla2017,Peng2019,felicetti2020universal,Zhu2020,Garbe2020}. 
Here,  the ``standard'' thermodynamic limit (diverging number of atoms) is replaced by a rescaling of the system parameters (diverging number of excitations). Finite-component phase transitions provide a tractable framework for theoretically assessing the properties of critical sensors~\cite{Chu2021,Garbe2022,Gietka2022,Gietka2022a,Garbe2022a,Ilias2022,Yang2022,ying_critical_2022,Salvia2023,DiFresco2024,Mihailescu_2024,alushi2024optimality} and offer a means to implement them with small-scale controllable devices~\cite{cai2021observation,DelteilNatMat19,Zejian2022,Fink2017,Brooks2021,chen2023quantum,Sett24}.

\begin{figure}
    \centering
    \includegraphics[width=1 \columnwidth]{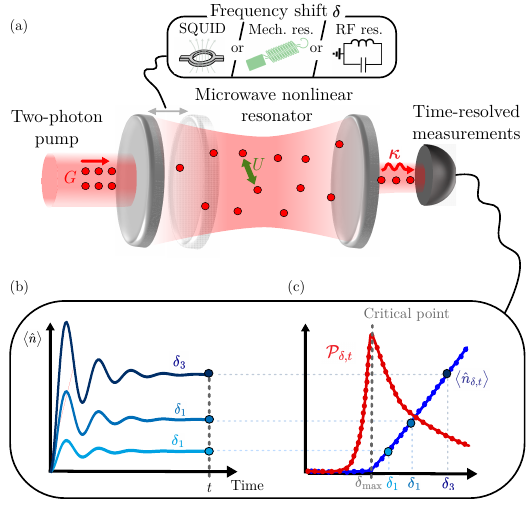}
    \caption{Sketch of the frequency estimation protocol. (a) A microwave resonator with Kerr nonlinearity $U$ is driven by a parametric (i.e. two-photon) pump $G$. The frequency of the cavity, and consequently the cavity-to-pump detuning $\delta$, changes when the (effective) cavity length is modified. This system is described by Eqs.~(\ref{Eq:Hamiltonian}) and (\ref{Eq:lindblad}). Such frequency tunability can be implemented, for example, using SQUIDs for magnetometry, optomechanical devices for force sensing, and RF resonators longitudinally coupled for MHz signal detection. (b) The cavity field escapes through photon loss at a rate $\kappa$ and is collected over time by a measurement apparatus, allowing for the reconstruction of the photon number in the cavity $\average{\hat{n}_{\detuning,t}}$. (c) From these time traces, the photon number at the steady state $\average{\hat{n}_{\detuning, {\rm ss}}}$ (blue curve) and the corresponding precision of the estimation of frequency $\mathcal{P}_{\delta,{\rm ss}}$ (red curve) are calculated. The maximal precision $\mathcal{P}_{\delta_{\rm max}, {\rm ss}}$ is achieved near the critical point of the second-order dissipative finite-component phase transition of the device.}
    \label{Fig:protocol schematic}
\end{figure}

Phase transitions~\cite{RotaPRL19,JinPRL13,LeePRL13,VicentiniPRA18,Foss-FeigPRA17} and their finite-components counterparts~\cite{CasteelsPRA16,BartoloPRA16,CarmichaelPRX15, CasteelsPRA17-2} can also occur in driven-dissipative settings \cite{Chen2023,FinkNatPhys18}, where the steady state of the system, rather than the ground state, undergoes a non-analytical change.
In particular, the parametrically (two-photon) driven-dissipative Kerr resonator exhibits first- and second-order finite-component dissipative phase transitions (DPT)~\cite{BartoloPRA16,MingantPRA18_Spectral}, both of which can be explored within a single parametrically-pumped 
resonator~\cite{beaulieuarxiv2023, PRXQuantum.4.020350}. This type of resonator plays a key role in superconducting quantum technology, such as in Josephson parametric amplifier \cite{yamamoto2008flux} and cat qubits \cite{putterman2024hardware,Rglade2024,PRXQuantum.4.020350}, both of which are build on systems nearly identical to the one considered here. 
Recent predictions suggest that criticality in the Kerr parametric oscillator could be used to implement optimal critical sensing protocols~\cite{Lorenzo2017,Ivanov2020,di2023critical,alushi2024optimality,Cabot24,alushi2024collective}, even in the presence of dissipation. For example, superconducting resonator operated near a first-order parametric criticality has been used in microwave photodetection~\cite{petrovnin2023microwave}.
Furthermore, since metrological experiments are often limited by a critical photons (or phonons) number---i.e., a maximal number of excitations that can be present in the sensor before unwanted effects emerge. As a result,
maximizing the amount of information extracted per photon becomes pivotal in such experiments, with a key example being the dispersive readout of a superconducting qubit \cite{Boissonneault2009}.

In this article, we demonstrate optimal precision scaling with a \emph{finite-component} critical quantum sensor at the steady state. We implement a frequency-estimation protocol with a two-photon driven nonlinear resonator [Fig.~\ref{Fig:protocol schematic}(a)]. Various platforms could be used to realize such system; in our case, the cavity is implemented with a superconducting resonator that is made both nonlinear and frequency-tunable by incorporating a SQUID.
We assess the metrological performance of the device by conducting time resolved measurements of the emission from the cavity [Fig.~\ref{Fig:protocol schematic}(b)] for various cavity-to-pump detuning. From these measurements, we show that the optimal precision on the frequency estimation is obtained near the critical point of the second-order driven-dissipative phase transition [Fig.~\ref{Fig:protocol schematic}(c)]. Additionally, by rescaling the system parameters, we demonstrate a quadratic scaling of the estimation precision with respect to the effective system size. We compare this result with the scaling of an optimal classical benchmark, which we theoretically prove to be at best linear with the system size. From a fundamental standpoint, the observed quadratic scaling is a direct signature of the quantum nature of the transition~\cite{montenegro2024}. From an applied perspective, this experiment serves as a proof of concept for the practical relevance of critical quantum sensing in solid-state quantum technologies.

\section{Device and Model} 

The device is a superconducting $\lambda/4$ cavity, made nonlinear by a SQUID shunting one end of the cavity to the ground~\cite{Note1}. 
A coherent tone, sent to a waveguide inductively coupled to the SQUID, generates an oscillating magnetic flux at nearly twice the cavity's resonance frequency~\cite{WilsonPRL10,Krantz_2013,lin2014josephson}. This modulation results in a two-photon, or parametric, drive. A coil placed beneath the sample is used to apply a constant flux bias, enabling the tuning of the SQUID's inductance. Tuning the inductance allows to change both the resonance frequency and the nonlinearity of the cavity. Further details can  be found in Ref.~\cite{beaulieuarxiv2023} and Appendix \ref{appendix:device}.

This system is described by the Hamiltonian \cite{beaulieuarxiv2023}
\begin{equation}
\label{Eq:Hamiltonian}
\hat{H}/\hbar = \detuning \hat{a}^\dagger \hat{a} + U/2 \, \hat{a}^\dagger \hat{a}^\dagger \hat{a} \hat{a} + G/2 \left( \hat{a}^\dagger\hat{a}^\dagger + \hat{a}\hat{a} \right) ,\end{equation} where $\hat{a}$ is the photon annihilation operator, $\detuning = \omega_r - \omega_p/2$ is the cavity-to-pump detuning
\footnote{Here, $\detuning$ includes all frequency drifts from nonlinear elements and coupling to external degrees of freedom, representing the detuning between the pump and measured frequency of the cavity, rather than the ``bare'' resonator frequency. Also notice the convention of calling the detuning $\omega$, as the symbol $\Delta$ is used to indicate errors.}, $U$ is the Kerr nonlinearity, and 
$G$ is the two-photon drive field amplitude.
Throughout the paper, we will study several values of $U$ and $G$ while sweeping $\detuning$.
The parametric signal emitted from the cavity is collected through a feedline coupled to the resonator on the opposite side of the SQUID, then filtered and amplified.
The interaction of the system with the feedline, fluxline, and other uncontrolled bath degrees of freedom, is captured by the Lindblad master equation
\begin{equation}\label{Eq:lindblad}
\begin{split}
\frac{\partial \rhot}{\partial t}   & = - \frac{i}{\hbar} [\hat{H}, \rhot] + \kappa 
 \frac{
2\hat{a} \rhot \hat{a}^\dagger - \{ \hat{a}^\dagger \hat{a}, \rhot\}}{2} ,
\end{split}    
\end{equation}
where the rate $\kappa$ is associated with the total photon loss. The total photon loss
$\kappa = \kappa_{\rm int} + 2 \kappa_{\rm ext}$ is the sum of the intrinsic losses $\kappa_{\rm int}$ and the coupling to each direction of the feedline $\kappa_{\rm ext}$, where we assumed the right- and left-coupling to be identical.
In the hanger and overcoupled configuration considered here, $2 \kappa_{\rm ext} \simeq \kappa$.
Other sources of noise are two-photon dissipation, dephasing, and thermal heating, which we find negligible compared to $\kappa/2\pi \simeq \SI{72}{\kilo \Hz}$~\cite{beaulieuarxiv2023} (c.f. Appendix \ref{appendix_section:device_parameters}).
As the system evolves, it will eventually reach its steady state $\sss$ defined by $\partial_t \sss =0$.

The system phases \cite{BartoloPRA16,MingantPRA18_Spectral,beaulieuarxiv2023,CalvaneseStrinati2024} and its sensing capability \cite{di2023critical} can be characterized by the average intracavity photon number and its variance. Experimentally, we do not have direct access to these quantities.  Instead, we monitor the output field of the cavity over time, after amplification with an effective gain $\mathcal{G}$. 
From input-output relations~\cite{walls_quantum_2008}, the measured output field is described by the bosonic output mode
$\hat{A}_{\detuning, t} = \sqrt{\mathcal{G}} [ \sqrt{\kappa_{\rm ext}} \hat{a}_{\detuning, t} + \hat{a}_{{\rm amp}, t}^\dag] $, where $\hat{a}_{\detuning, t}$ is the intracavity field $\hat{a}$ in the Heisenberg picture and $ \hat{a}_{{\rm amp}, t}$ is the amplifier noise mode.
The output power is then $\hat{N}_{\detuning, t}= \hat{A}^\dagger_{\detuning, t} \hat{A}_{\detuning, t}$.
As the amplifier noise is uncorrelated to the photon emission, and it averages to zero, the expectation value  $\langle \hat{N}_{\detuning, t}\rangle
= \mathcal{G} [\langle \hat{n}_{\detuning,t}\rangle + n_{\rm amp}] $, where
$\average{\hat{n}_{\detuning, t}} = \kappa_{\rm ext} \operatorname{Tr}[\hat{a}^\dagger \hat{a} \rhot_{\delta,t}]$ is proportional to the intracavity population and $ n_{\rm amp}$ is the contribution of the amplifier~\cite{Silva10,Eichler2011,di2014dual} (see Appendix \ref{appendix_section:precision_estimation}).
Experimentally,  $\average{\hat{n}_{\detuning, t}}$, can be obtained by substracting the amplified noise $\mathcal{G} n_{\rm amp}$ (measurement background) from $\langle \hat{N}_{\detuning, t}\rangle$. As discussed below, the calibration of gain is not necessary for the metrological protocol discussed here. Therefore, $\average{\hat{n}_{\detuning,t}} $ will always be given in arbitrary units.

As we are investigating the metrological properties of the steady state~\cite{di2023critical}, most of the measurements presented below focus on the steady-state properties (i.e. $t={\rm ss})$. 
Consequently, we measure $\langle \hat{N}_{\detuning, t}\rangle$ for $t \gg 1/\kappa$ at discrete time intervals. To accumulate statistics, we collect all the  data from $t>\SI{15}{\micro\second} \simeq 7 /\kappa$ up to an arbitrary long time, fixed in the experiments at $T=\SI{69}{\micro\second}$ (see the Appendix \ref{appendix_section:precision_estimation} for details).
We then average over $4\times 10^5$ or $8\times 10^5$  independent time traces, depending on the acquisition rate of $6.7\times 10^{5}$ samples per second or $4\times 10^{6}$ samples per second to obtain $\average{\hat{N}_{\detuning, t}}$ .

\section{Scaling towards the thermodynamic limit}

\begin{figure}
    \centering
    \includegraphics[width=1 \columnwidth]{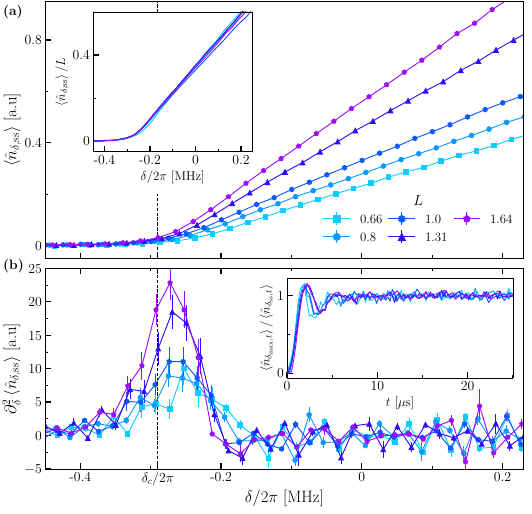}
    \caption{Signature of the second-order DPT when scaling towards the thermodynamic limit using Scaling (I) in Eq.~\eqref{Eq:Scaling1}. (a) The expectation value $\average{\hat n_{\detuning,{\rm ss}}} \propto \operatorname{Tr}[\rho_{\delta, {\rm ss}} \hat{a}^\dagger \hat{a}]$ for various $L$ as a function of the detuning $\detuning$.  The expectation value is obtained by averaging over $M=4\times 10^5$ time traces sampled with an acquisition rate of $6.7 \times 10^{5}$ samples per second. The error bars correspond to $\sqrt{\Delta n_{\delta,ss}^2/M}$  where $\Delta n_{\delta,{\rm ss}}$ is the standard deviation over the $M$ measurements.
    The inset shows the rescaled photon number $\average{\hat n_{\detuning,{\rm ss}}}/L$, demonstrating that all curves collapse onto each other. These data have been obtained by subtracting the amplifier noise $n_{\rm amp}$, as detailed in Appendix \ref{appendix_section:precision_estimation}.
    (b) Finite-difference estimation of the second-order derivative $\partial_{\detuning}^2 \average{\hat n_{\detuning,{\rm ss}}}$. The error bars are calcualted using error-propagation applied to the finite-difference formula. The inset shows the averaged normalized time trace $\average{\hat n_{\detuning_{\rm max},t}}$, where $\detuning_{\rm max}$ corresponds to the peak of  $\partial_{\detuning}^2 \average{\hat n_{\detuning,{\rm ss}}}$. To capture the fast dynamics, the acquisition rate was set to $4 \times 10^{6}$ samples per second for the measurement and the number of averaged traces increased to $M=8\times 10^5$. In both figures, the vertical dashed line marks the expected detuning at which the second-order DPT  occurs, given by $\detuning_c \equiv - \sqrt{G^2 -\kappa^2}$.  Parameters: $G/2\pi \simeq \SI{300}{\kilo \Hz}$, $\tilde{U}/2\pi = \SI{-9.14}{\kilo \Hz}$ [where $U = \tilde{U}/L$] and $\kappa /2\pi \simeq \SI{72}{\kilo \Hz}$ (details in Appendix \ref{appendix:scaling}).}
    \label{fig:second_order_DPT}
\end{figure}

In Fig.~\ref{fig:second_order_DPT}(a), we show $\average{\hat n_{\detuning,{\rm ss}}}$ as a function of detuning $\detuning$ for $G/2\pi \approx \SI{300}{\kilo \Hz}$ and for various values of $U$ obtained by tuning the DC flux bias (see the Appendix \ref{appendix:scaling}). 
All curves show a transition from (i) a vacuum-like state at
$\delta\ll 0$ 
to
(ii) a bright state with a linearly growing number of photons. The passage between (i) and (ii) occurs approximately at $\detuning_c \equiv - \sqrt{G^2 -\kappa^2}$ and is indicated by the vertical dashed line in the figures.

The finite-component system described by Eq.~\eqref{Eq:lindblad} is known to display a second-order DPT in the thermodynamic limit \cite{beaulieuarxiv2023,PRXQuantum.4.020350}. In this limit, the transition between phases (i) and (ii) remains continuous but becomes non-differentiable. The thermodynamic limit can be reached by one of the following scalings~\cite{BartoloPRA16}

\begin{align}
\label{Eq:Scaling1}
    & (\text{I}): &  \!\!\!  \detuning &= \tilde{\detuning} &   G&= \tilde{G} &   U&=  \tilde{U}/L  &  \kappa &= \tilde{\kappa}, \\
    \label{Eq:Scaling2}
     & (\text{II}): &  \!\!\!  \detuning&= \tilde{\detuning} L  &  G&=   \tilde{G} L  &   U&= \tilde{U}  &   \kappa&= \tilde{\kappa}L,
\end{align}
where the $\tilde{v}$ indicates the scaled variable associated with the physical variable $v$.
Here, $L$ is a number that plays a role similar to the number of atoms or the lattice size in an extensive system.
Scaling towards the thermodynamic limit means considering increasing values of $L$ and comparing curves with the same $\tilde{v}$ variables.
In the rescaled pictures, curves of $\average{\hat n_{\detuning, {\rm ss}}}/L$ for various $L$ collapse onto each other. This is shown for the Scaling (I) in the inset of Fig.~\ref{fig:second_order_DPT}(a). Slight deviations from perfect rescaling at larger detuning values are attributed to experimental errors in the photon number, such as inaccuracies in the calibration of the two-photon drive $G$ (see Appendix \ref{appendix:scaling}) and the frequency-dependent attenuation or gain of microwave components.The emergence of the second-order DPT can be seen by calculating $\partial^2_{\detuning}  \average{\hat n_{\detuning, {\rm ss}}}$ and comparing curves with the same re-scaled parameters while increasing $L$. 
As shown in Fig.~\ref{fig:second_order_DPT}(b), $\partial^2_{\detuning} \average{\hat n_{\detuning, {\rm ss}}}$ has a maximum whose value grows with $L$, confirming the expected onset of non-analiticity in the thermodynamic limit.\\
Phase transitions are also characterized by critical slowing down, i.e., the time required for the system to evolve becomes divergently large as the system approaches the thermodynamic limit.
In open quantum systems, critical slowing down manifests differently depending on the observables. 
For instance, observables associated with symmetry breaking at the critical point (such as $\hat{a}$ in our system) slow down exponentially as $L$ increases. This was experimentally shown in Ref.~\cite{beaulieuarxiv2023}, with a law $\langle\hat{a}_t \rangle \propto e^{-\alpha L t}$ with $\alpha$ a constant of the model.  Observables with different symmetry features can, however, follow different dynamics. Specifically, for our system and metrological protocol, the observable of interest is $\hat{a}^\dagger \hat{a}$, which is \emph{not} the order parameter associated with spontaneous symmetry breaking.
In the inset of Fig.~\ref{fig:second_order_DPT}(b), we plot the normalized $\average{\hat n_{\detuning_{\rm max},{\rm ss}}}$ as a function of time and for various $L$.
Despite scaling towards the thermodynamic limit, we observe a negligible difference in the time required to reach the steady state for curves with different $L$.

In the following, we focus on the task of measuring a small frequency change $\detuning \to \detuning + \discretization$. The change of $\detuning$ can be achieved by varying either the resonator frequency $\omega_r$ or the pump frequency $\omega_p$.  Since this work aims at studying the sensing properties, we chose to vary the pump frequency $\omega_p$, as this allows for precise and highly controlled adjustments. However, for practical applications, variations of $\delta$ could also be induced by changes of $\omega_r$. We will show that the Scaling (I) in Eq.~\eqref{Eq:Scaling1} leads to a critically-enhanced scaling of the frequency-estimation precision. In practice, the maximum achievable value of $L$ is limited by the Kerr nonlinearity. Therefore, we will use the Scaling (II) in Eq.~\eqref{Eq:Scaling2} as a proof-of-concept demonstration for  precision scaling with larger values of $L$, keeping in mind that this also implies a scaling of the estimated parameter itself.
\begin{figure}
    \centering
    \includegraphics[width=1 \columnwidth]{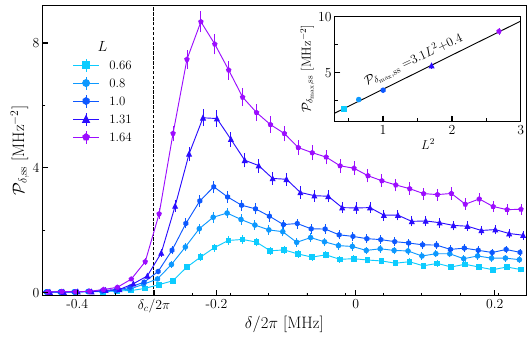}
    \caption{ Quadratic dependence of the maximal precision $\mathcal{P}_{\detuning_{\rm max}, {\rm ss}}$ on $L$. Precision of the estimation $\mathcal{P}_{\detuning, {\rm ss}}$ for the Scaling (I) in Eq.~\eqref{Eq:Scaling1} calculated as defined in Eq.~(\ref{Eq:error}). The error bars are calcualted using error-propagation applied to the Eqs.~(\ref{Eq:error}) and (\ref{eq:precision_definition}) (see Appendix \ref{appendix_section:precision_estimation} for details).
    The inset shows the maximal precision $\mathcal{P}_{\detuning_{\rm max}, {\rm ss}}$ as a function of $L$. The grey line is a fit of the data, demonstrating a quadratic dependence of $\mathcal{P}_{\detuning_{\rm max}, {\rm ss}}$ on $L$. Parameters as in Fig.~\ref{fig:second_order_DPT}(a).
    }
    \label{fig:scaling_I_enhancement}
\end{figure}

\section{Enhanced sensing} 

Given a measurement of  $\hat{N}_{\detuning, t}$, the error in the estimation of $\detuning$ depends on $\average{\hat{N}_{\detuning, t}}$ and on the standard deviation $\standarddeviation{{N}_{\detuning, t}}= \sqrt{\average{\hat{N}_{\detuning, t}^2} - \average{\hat{N}_{\detuning, t}}^2}$. The error in the estimation can be characterized by the error propagation formula~\cite{paris_quantum_2009}
\begin{equation}\label{Eq:error}
     \error{\detuning}_{\rm ss}  = \frac{\standarddeviation{N_{\detuning,{\rm ss}}}}{\left|\partial_{\detuning} 
    \average{\hat{N}_{\detuning, {\rm ss}}} \right|} \simeq 
    \frac{\left(\standarddeviation{N_{\detuning,{\rm ss}}} + \standarddeviation{N_{\detuning+\discretization,{\rm ss}}}\right) \discretization}{2 \left|\average{\hat{N}_{\detuning, {\rm ss}}}- \average{\hat{N}_{\detuning + \discretization, {\rm ss}}}\right| },
\end{equation}
where the last part represents a discretized version of the error.

The choice of $\hat{N}_{\delta,\rm ss}$ implies that the intracavity photon number $\average{\hat{n}_{\detuning,{\rm ss}}}$ is the resource used to estimate $\detuning$.
Therefore, we expect that the error in the estimation of $\detuning$ will decrease as the photon number increases. Furthermore, Eq.~\eqref{Eq:error} shows that the precision in the estimation of $\detuning$ is independent of any proportionality factor applied to $\hat{N}_{\delta,\rm ss}$, making a calibration of the gain value $\mathcal{G}$ unnecessary. 
Note, however, that the error is \textit{not} independent on $n_{\rm amp}$
and increases as $ n_{\rm amp}$ does.
Since $\average{\hat n_{\detuning, {\rm ss}}} \propto L$ [see  the inset of Fig.~\ref{fig:second_order_DPT}(a)], we introduce the precision of the estimation $\mathcal{P}_{\detuning, {\rm ss}}$ of $\detuning$ 
\begin{equation}
\label{eq:precision_definition}
     \mathcal{P}_{\detuning, {\rm ss}}=\left( \error{\detuning _{\rm ss} } \right)^{-2}\propto  L^{\beta}.
\end{equation}
where $\beta~=~1~[=~2]$ is an upper bound for a classical [quantum] resource~\cite{alushi2024optimality} (see Appendix \ref{appendix:classical}).
In Fig.~\ref{fig:scaling_I_enhancement}, we plot $\mathcal{P}_{\detuning, {\rm ss}}$  for various values of $L$.
We remark that by increasing $L$, the estimation becomes more precise.
Each curve displays a maximum, occurring at $\detuning_{\rm max}$.
In the inset, we observe a quadratic dependence between the maximum of the precision and $L$.

\subsection*{Connecting criticality and enhanced sensing}
In the device used for this study, the Kerr nonlinearity can be tuned via the flux bias between $ \SI{-13.86}{\kilo \Hz} \leq U/{2\pi} \leq \SI{-5.58}{\kilo \Hz}$, corresponding to $0.66\leq L \leq 1.64$.
As anticipated above, to explore larger values of $L$ we resort to the Scaling (II) in Eq.~\eqref{Eq:Scaling2}.
Although we cannot independently scale $\kappa$ without affecting $U$, both $\detuning$ and $G$ can be controlled by changing the frequency and amplitude of the parametric drive, respectively.

We recall here that, given a set of reduced parameters $\{\tilde{\detuning}, \tilde{G}, \tilde{U}, \tilde{\kappa}\}$ and a given value of $L$, the two scalings in Eqs.~\eqref{Eq:Scaling1} and \eqref{Eq:Scaling2} lead to different physical parameters. 
It is, however, meaningful to compare the results for the same rescaled parameters.
Namely, we should compare $\average{\hat n_{\detuning,{\rm ss}}}$ vs $\detuning$ for Scaling (I) and  $\average{\hat n_{\tilde{\detuning},{\rm ss}}}$ vs $\tilde{\detuning}$ for Scaling (II). Similarly, $\mathcal{P}_{\tilde{\detuning}, \rm{ss}} = \mathcal{P}_{{\detuning}, \rm{ss}}$ for Scaling (I) and $\mathcal{P}_{\tilde{\detuning}, \rm{ss}} = \mathcal{P}_{\detuning/L, {\rm ss}}$ for Scaling (II).
In supplementary Fig.~\ref{fig:scaling_proof}, we verify that the two scalings lead to similar results both for the photon number and the precision upon the appropriate rescaling . The overlap of curves with identical values of $L$, but different scaling indicates that the Scaling (II) can be used to qualitatively explore larger values of $L$ that cannot be reached in our device using the Scaling (I).

\begin{figure}
    \centering
    \includegraphics[width=1 \columnwidth]{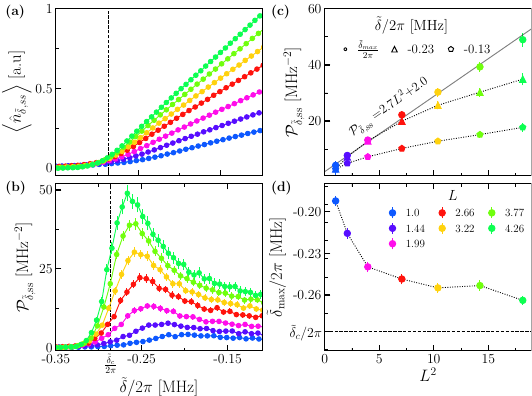}
    \caption{For the Scaling II in Eq.~\eqref{Eq:Scaling2}: (a) Output photon number at the steady state $\average{\hat n_{\tilde{\detuning}, {\rm ss}}}$ vs scaled detuning $\tilde{\detuning}$ for increasing $L$. (b) Precision  $\mathcal{P}_{\tilde{\detuning}, {\rm ss} }$ as a function of $\tilde{\detuning}$.  The scaling of $\mathcal{P}_{\tilde{\detuning},  \rm{ss}}$ as a function of $L$ for various points, including the optimal point $\tilde{\detuning}_{\rm max}$ for each $L$. (c) $\mathcal{P}_{\tilde{\detuning}_i, {\rm ss}}$ for $\tilde{\delta_i}= \tilde{\delta}_{\rm max}\text{, } \tilde{\delta_i}/2\pi = -\SI{0.23}{\mega\hertz} \text{ and } \tilde{\delta_i}/2\pi=-\SI{0.13}{\mega\hertz}$. The solid grey line is a fit of the data demonstrating quadratic dependence for $\mathcal{P}_{\tilde{\detuning}_{\rm max}, {\rm ss}}$, while the two light grey dotted lines linking the markers are included as visual guide. 
    (d) $\tilde{\detuning}_{\rm max}$ as a function of $L$. The light grey dotted line linking the markers is included as visual guide. Error bars are calculated as in Figs.~\ref{fig:second_order_DPT}(a) and \ref{fig:scaling_I_enhancement}. The error on  in panel (c) corresponds to the size of a detuning step.}
    \label{fig:scalingII}
\end{figure}

Therefore, we analyze the metrological properties of the device for larger values of $L$ for the Scaling (II).
In Fig.~\ref{fig:scalingII}(a), we plot the output photon number at the steady state, $\langle{\hat n_{\tilde{\detuning},{\rm ss}}}\rangle$, observing the characteristics indicative of the onset of a second-order DPT. 
In Fig.~\ref{fig:scalingII}(b) we plot $\mathcal{P}_{\tilde{\detuning}, {\rm ss}}$ and confirm that the system gains in precision as it scales towards the thermodynamic limit.
Figure \ref{fig:scalingII}(c) shows the maximum of $\mathcal{P}_{\tilde{\detuning}_{\rm max}, {\rm ss}}$ as function of $L$; the data are in line with a quadratic scaling, comparable to that observed in Fig.~\ref{fig:scaling_I_enhancement}. 
In the same panel, we also plot $\mathcal{P}_{\tilde{\detuning}_i, {\rm ss}}$ for $\tilde{\delta_i}/2\pi=\SI{-0.23}{\mega\hertz} \text{ and } \tilde{\delta_i}/2\pi=-\SI{0.13}{\mega\hertz}$, highlighting the fact that the optimal scaling is achieved around the optimal point $\tilde{\detuning}_{\rm max}$.
Finally, in Fig.~\ref{fig:scalingII}(d) we show that as $L$ increases, the difference between the critical point $\tilde{\detuning}_c$ and the point of maximal precision $\tilde{\detuning}_{\rm max}$ decreases, suggesting that the two will eventually coincide for large enough $L$. These data show the enhanced sensing capabilities of the parametrically driven Kerr resonator, and that this enhanced sensing occurs near the critical point.

\section{Discussion and conclusion}  
In classical pump-and-probe experiments, doubling the pump power does not result in twice the precision.
More formally, as shown in the Appendix \ref{appendix:classical}, it can be demonstrated that a protocol based on a linear resonator driven by a coherent drive achieves a maximal precision bounded by $\mathcal{P}_{\delta_{\rm max}, {\rm ss}} \propto \langle \hat{a}^\dagger \hat{a}\rangle$, the number of photon in the resonator, even when optimizing over all system and drive parameters. This remains true even in the absence of any noise,  internal dissipation or decoherence. 

Our experiment is still pump-and-probe, but we have observed a quadratic scaling of the parameter-estimation precision, $\mathcal{P}_{\delta_{\rm max}, {\rm ss}} \propto L^2 \propto \langle \hat{a}^\dagger \hat{a}\rangle^2$.
The key difference is that the system is operated in the vicinity of the critical point of a second-order dissipative phase transition.
The system's nonlinearity and the parametric quantum process that converts the external drive into a two-photon pump make it possible to overcome the classical precision bound derived in \ref{appendix:classical}. Overcoming this bound and maximizing the information per photon is essential in many metrological experiments limited by a critical photon number, such as in the dispersive readout of a qubit \cite{Boissonneault2009}. From a fundamental perspective, our experiment demonstrates that quantum sensing protocols are a valuable tool for characterizing the quantum nature of driven-dissipative phase transitions. Technologically, our results pave the way to the development of a new generation of optimal quantum sensors~\cite{alushi2024optimality} based on solid-state critical systems. As sketched in Fig.~\ref{Fig:protocol schematic}, our sensing protocol can be used to measure various physical quantities by choosing an appropriate dispersively coupled component. In the context of superconducting circuits operating at cryogenic temperatures, relevant examples include detecting magnetic fields with a superconducting quantum inderference device (SQUID)~\cite{halbertal_imaging_2017}, forces with optomechanical devices~\cite{barzanjeh_optomechanics_2022} and MHz signals with longitudinally coupled RF resonators~\cite{bothner_photon-pressure_2021,rodrigues_parametrically_2022,Rodriguez24}. 
Finally, this experimental demonstration might foster the implementation of critical sensing protocols in different  contexts, such as atomic physics~\cite{Puebla2017,cai2021observation} or nanoelectronics~\cite{Mihailescu_2024}.



\begin{acknowledgments}
The authors thank R. Puig I Valls, L. Peyruchat, A. Mercurio, and V. Savona for useful discussions, and V. Jouanny, F. Oppliger and F. De Palma for helping with the measurement setup.


P.S. acknowledges support from the Swiss National Science Foundation (SNSF) through the Grant Ref. No. 200021 200418, Grant Ref. No. 206021\_205335, and from the Swiss State Secretariat for Education, Research and Innovation (SERI) through grant 101042765 SEFRI MB22.00081. P.S., F.M. and S.Fr. acknowledge support from the EPFL Science Seed Fund 2021. P.S. and G.B. acknowledge support from the Swiss National Science Foundation through the Grant Ref. No. UeM019-16 - 215928. M.S. acknowledges support from the EPFL Center for Quantum Science and Engineering postdoctoral fellowship. S.Fe. acknowledges financial support from PNRR MUR project PE0000023-NQSTI financed by the European Union – Next Generation EU. R.D. acknowledges support from the Academy of Finland, grants no. 353832 and 349199.
\end{acknowledgments}

\appendix


\newcommand{\IN}[1]{\hat{#1}_{\rm in}}
\newcommand{\OUT}[1]{\hat{#1}_{\rm out}}

\section{Device and setup}

\label{appendix:device}

\subsection{Design and fabrication}
\label{appendix_section:device}

\begin{figure}[!htb]
        \centering
        \includegraphics[width=\columnwidth]{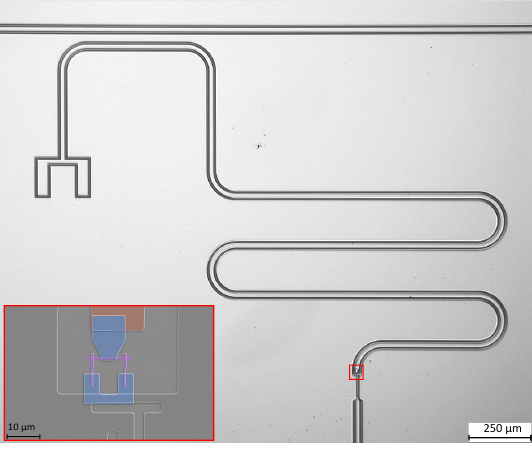}
        \caption{Optical and SEM images of the device. Micrograph of the device: a $\lambda/4$ resonator terminated by a SQUID on one end and capacitively coupled to a feedline on the other. In the lower left corner, an SEM of the SQUID, formed by two Josephson junctions (purple), is shown. The SQUID is galvanically connected to the resonator (red) by a patch (blue). The upper part of the fluxline is visible beneath the lower patch. The zoomed-in area corresponds to the red rectangle in the micrograph. 
        }
         \label{Fig:device}
\end{figure}

The device consists of a 150 nm thick aluminum layer on a 525 µm thick high-resistivity silicon substrate. It features a coplanar waveguide resonator fabricated by photolithography followed by wet etching. The nonlinear resonator is grounded through two Al/AlO$_{\rm x}$/Al Josephson junctions forming a superconducting quantum interference device (SQUID). The junctions are made by e-beam lithography using a double-evaporation technique. Further details on the fabrication can be found in Ref.~\cite{beaulieuarxiv2023}. Figure \ref{Fig:device} shows an optical micrograph of the sample used in the experiment, with an SEM zoom of the SQUID in the lower left corner. 

\subsection{Experimental setup}

\begin{figure*}[h]
        \centering
        \includegraphics[width=\textwidth]{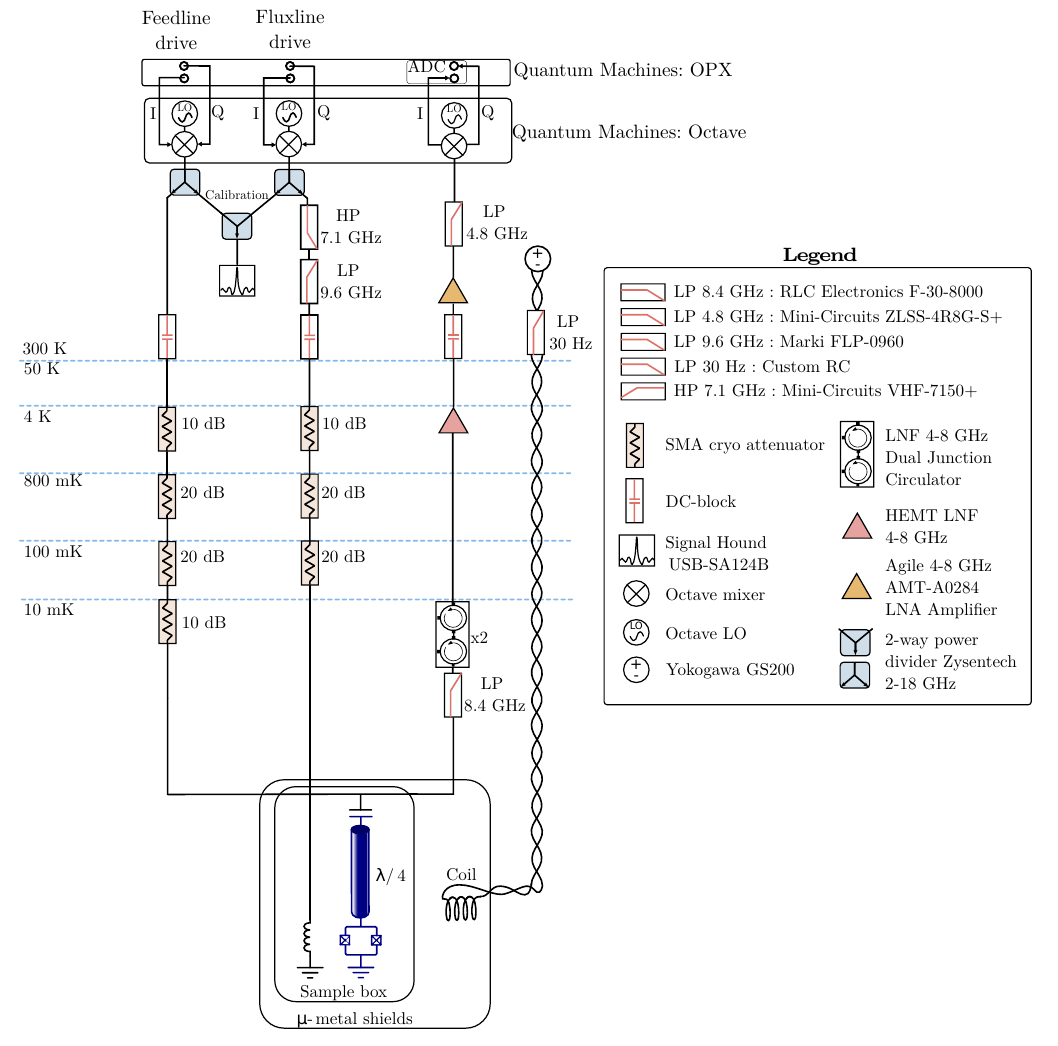}
        \caption{Schematic of the experimental setup. A two-photon drive is generated by applying a signal to the flux line, which modulates the magnetic flux in the SQUID at frequency $\omega_p \sim 2\omega_r$. This parametric excitation of the cavity leads to emission of a signal at $\omega_p/2$. The emitted signal is then amplified and filtered to eliminate any componet at $\omega_p$. The output signal is downconverted to an intermediate frequency to acquire the two quadratures $I$ and $Q$. These intermediate frequency signals are then digitaly demodulated and integrated within the OPX + over a time interval $\Delta T$, producing a single pair of $I$ and $Q$ values.  }
         \label{Fig:experimental set up}
\end{figure*}

The packaged sample is mounted in a high-purity copper enclosure, thermally anchored at the mixing chamber stage of a BlueFors dilution refrigerator.  A NbTi coil placed beneath the sample is connected to a current source Yokogawa GS200 to generate a DC flux bias.  Two high permeability metal cans provide shielding against external magnetic fields. The cryogenic and room temperature setup is depicted in Fig.~\ref{Fig:experimental set up}. An OPX+ and Octave modules generate drives at $\sim \omega_r$ (single-photon drive for device characterization) and $\sim 2\omega_r$ (the one which gets converted by three-wave mixing into the two-photon drive). The flux line has 10 dB, 20 dB, and 20 dB attenuators positioned at the 4 K, 800 mK, and 100 mK stages, respectively. The single-photon drive line includes an extra 10 dB attenuator at the mixing chamber stage. Throughout the experiment, the local oscillator used by the Octave to upconvert the signal and used for the single-photon drive remains off to prevent any leakage field in the feedline. It is turned on only before/after the measurements are completed to extract the device parameters. After exiting the Octave, the two pulses are split using a 2-way power divider ZSPD-20180-2S. Half of the signal enters the fridge, while the other half is directed to a spectrum analyzer (Signal Hound USB-SA124B) to monitor the drive amplitude and compensate for any drift. The output signal, collected via the feedline, passes through two circulators (LNF 4-8 GHz Dual Junction Circulator) and travels in a NbTi low-loss superconducting line before being amplified by a 4-8 GHZ LNF High-Electron-Mobility Transistor (HEMT) at the 4K plate. The output signal is further amplified at room temperature using a low noise amplifier (Agile AMT-A0284) before being demodulated in the Octave and digitized in the OPX+.

\subsection{Characterization of the device parameters}
\label{appendix_section:device_parameters}

Circuit quantization of a distributed $\lambda/4$ resonator terminated by a SQUID allows us to calculate key parameters of the system, i.e. $\omega_r$ and $U$, based on the distributed circuit characteristics. The complete quantization procedure is detailed in Refs. \cite{beaulieuarxiv2023,MWallquist_strip, Eichler_2014_josephson}. Here, we summarize some of the key results. 

\begin{figure}[h]
        \centering
        \includegraphics[width=\columnwidth]{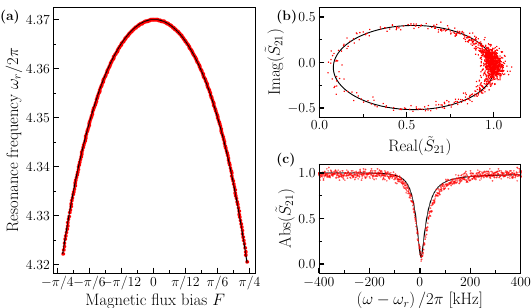}
        \caption{ Measurement of the device parameters. (a) Resonance frequency $\omega_r$ for varying magnetic flux bias $F$. (b) Real against imaginary part of the measured normalized scattering coefficient   $\tilde{S}_{21}= S_{21}/a e^{j\alpha}e^{-2 \pi i f \tau}$ . (c) The absolute value of the measured normalized scattering coefficient. In all panels, the red markers are the measurements, while the black line indicates the fit to Eq.~(\ref{eq:flux_response_approx}) for panel (a) and Eq.~(\ref{supp_eq: scattering_response}) for panels (b) and (c). } 
        \label{Fig:parameter_char}
\end{figure}

The wavevector of the fundamental mode $k_0$ can be calculated from the transcendental equation 
\begin{equation}\label{Eq:eigenmode}
    \frac{L_{cav}}{L_J}-\frac{C_J(k_0 d)^2}{C_{cav}}=k_0 d \tan(k_0d), 
\end{equation}
where $L_{cav}$ is the inductance of the resonator, $L_J$ and $C_J$ are respectively the inductance and capacitance of the SQUID and $d$ is the length of the resonator. Expressed in terms of resonance frequency and in the limit of small participation ratio $\gamma=\frac{L_J}{L_{\rm cav}}$, Eq.~(\ref{Eq:eigenmode}) can be rewritten as
\begin{equation}\label{eq:flux_response_approx}
    \omega_0 \approx \frac{\omega_{\lambda/4}}{1+\gamma},
\end{equation}
where $\omega_{\lambda/4}$ is the resonance frequency of the fundamental mode of the bare cavity for $L_J=0$. Note that $\gamma$ varies with the external flux, since the SQUID indutance changes according to $L_J=\frac{L_{J}(F=0)}{|\cos(F)|}$, where $F=\pi \Phi_{\rm dc}/\Phi_0$, with $\Phi_{\rm dc}$ being the DC flux bias and $\Phi_0$ the flux quantum. We use Eq.~(\ref{eq:flux_response_approx}) to fit the measured resonance frequency as a function of flux, which allows us to extract $\gamma$ and $\omega_{\lambda/4}$. The magnetic flux is varied by adjusting the current sent to the coil underneath the sample. The result of this measurement is shown by the red datapoints in Fig.~\ref{Fig:parameter_char}(a) and the fit by the black line. From the fit, we find the participation ratio $\gamma=\SI{3.1e-2}{}$ and the bare cavity frequency $\omega_{\lambda/4}/2\pi =\SI{4.5068}{\giga\Hz}$ . Knowing $\gamma$ and $\omega_{\lambda/4}$, we can calculate the Kerr nonlinearity  at different flux points using 
\begin{equation}\label{eq:Kerr}
      U  = -\frac{\hbar \omega_0^2 L_{cav}}{2\gamma \phi_0^2}  \left [ \frac{ \cos^2(k_0d)}{(k_0d)^2 M_0} \right ] ^2,
\end{equation}
where, 
\begin{equation}
    M_0 = \left [ 1+ \frac{\sin(2 k_0 d)}{2k_0d} + \frac{2C_J}{C_{cav}}\cos^2(k_0d) \right ].
\end{equation}
The value of $F$ and $U$ for the different operating points are reported in Tables.~\ref{table:scaling_I} and \ref{table:scaling_II} .\\

Futhermore, for each operating point, the resonance frequency $\omega_r$ and the coupling to the environment $\kappa$ are determined by fiting the scattering response $S_{21}$. The scattering response is measured by spectroscopy using the feedline (see Fig.~\ref{Fig:experimental set up}). For small probe signal amplitude, the scattering coefficient of the hanger resonator can be expressed as \cite{Probst_2015}
\begin{equation}
\label{supp_eq: scattering_response}
  S_{21}(f) = a e^{i\alpha} e^{-2\pi i f \tau} \left[ 1 - \frac{(Q_l/\left| Q_c \right|)e^{i\phi}}{1+2iQ_l(f/f_r-1)} \right], 
\end{equation}
where $f$ is the probe frequency, $f_r$ is the resonance frequency, $Q_l$ and  $\left| Q_c \right|$ are respectively the loaded and absolute value of the coupling quality factor, and $\phi$ quantifies the impedance mismatch. The fit is performed using the algorithm provided in \cite{Probst_2015}. The red markers in Fig.~\ref{Fig:parameter_char}(b) show the measured real and imaginary parts of the normalized scattering response $\tilde{S}_{21}= S_{21}/a e^{j\alpha}e^{-2 \pi i f \tau}$, while the black line corresponds to the fit. From the fit, we obtain resonance frequency $\omega_r=2\pi f_r$ and the external coupling $\kappa = \omega_r/Q_l$. The values are also reported in Table.~(\ref{table:scaling_I}). \\

The final parameter to be estimed is the amplitude of the two-photon pump $G$ at the device. To estimate $G$, we use the measurement of the steady state photon number $\average{\hat n_{\detuning, \rm ss}}$ as a function of the detuning $\detuning$. In the mean-field approximation, the stable solution for $\average{\hat n_{ \rm ss,\detuning}}$ is given by 
\begin{equation}
\label{eq:steady_state_photon_number}
    \average{\hat n_{ \detuning,\rm ss}} \approx \frac{\detuning+\sqrt{\left | G \right |^2-\kappa^2}}{|U|}
\end{equation}
Note that Eq.~(\ref{eq:steady_state_photon_number}) assumes a two-photon dissipation rate $\kappa_2$ and dephasing rate $\kappa_{\phi}$ that are  much smaller than $\kappa$. This assumption holds in our case. In \cite{beaulieuarxiv2023}, we have estimated the different rates to be of the order $\kappa_{\phi}/2\pi \approx \SI{4}{\kilo\hertz}$, $\kappa_2/2\pi \approx\SI{80}{\hertz}$ and $\kappa/2\pi \approx \SI{70}{\kilo\hertz}$.  In this regime, we can find the value of $G$ by extrapolating  $\average{\hat n_{\detuning,\rm ss}}$ to the x-intercept, denoted as $\detuning_{0}$ $[\average{\hat n_{\delta_0, \rm ss}}=0]$
even without knowing the Kerr nonlinearity. The value of $G$ is then given by 
\begin{equation}
\label{eq:extrapolation G}
    G= \sqrt{\detuning_{0}^2+\kappa^2}
\end{equation}

\begin{figure}[h]
        \centering
        \includegraphics[width=\columnwidth]{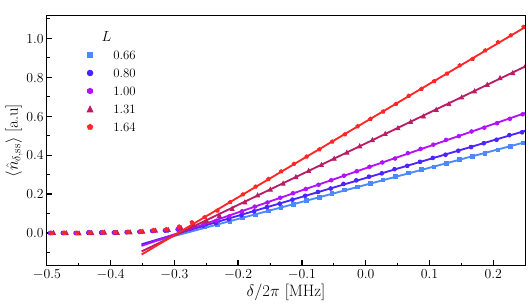}
        \caption{ Estimation of the two-photon pump $G$. Markers indicate the expectation value $\average{\hat n_{ \detuning, \rm ss}}$ for the various $L$ as a function of the detuning $\detuning$. The solid lines corresponds to a linear fit. Extrapolation of the linear fit to $\average{\hat n_{\detuning_{0},\rm ss}}=0$ allows to estimate the value of $G$ using Eq.~(\ref{eq:extrapolation G}). } 
        \label{Fig:G_estimation}
\end{figure}

In Fig.~\ref{Fig:G_estimation}, we show the measurement of $\average{\hat{n}_{\rm ss, \detuning}}$ as a function of $\detuning$ for Scaling (I) (different values of $U$). The experimental data are shown by markers while the solid lines correspond to a linear fit. By extrapolating the fit to $\average{\hat{n}_{\detuning_{0},\rm ss}}=0$, we observe that all curves cross at a detuning $\detuning_0$, indicating that the value of $G/2\pi=\SI{300}{\kilo \hertz}$ is consistent across all curves.

\section{Implementation of the scalings and precision calculation}
\label{appendix:scaling}

As detailed in the main text, the Scaling (I) is charactherized by the following transformations : 
\begin{align}
\label{Eq:Scaling1_supp}
    & (\text{I}): &  \!\!\!  \detuning &= \tilde{\detuning} &   G&= \tilde{G} &   U&=  \tilde{U}/L  &  \kappa &= \tilde{\kappa}. 
\end{align}
To perform this scaling, it is necessary to change only the value of $U$ while keeping all other parameters fixed. The value of $U$ is adjusted by varying the external flux $F$ as described by Eq.~(\ref{eq:Kerr}). Changing the value of $F$ will also change the value of $\omega_r$ as described by Eq.~(\ref{eq:flux_response_approx}). This can be compensated by modifying the pump frequency  $\omega_p$ to maintain a constant detuning. Table \ref{table:scaling_I} summarizes the different operating points and the parameters used to realize the Scaling (I). In this table, the value of $\omega_r$ and $\kappa$ are obtained by fitting the scattering coefficient $S_{21}$. The value of $U$ and $F$ are derived from Eq.~(\ref{eq:flux_response_approx}) and Eq.~(\ref{eq:Kerr}), respectively. The value of $G$ is determined by extrapolating $\average{\hat{n}_{\detuning, \rm ss}}$ to the x-intercept as described by Eq.~(\ref{eq:extrapolation G}). The pump frequency is set experimentally, and the detuning is defined as $\detuning=\omega_r-\omega_p/2$. Finally, the value of $L$ is simply defined from the scaling on the Kerr nonlinearity $U$. 
\begin{table*}
\centering
\caption{\textbf{Operating points and parameters used for Scaling (I)}}
\begin{tabular}{>{\centering\arraybackslash} p{1.25cm}
>{\centering\arraybackslash} p{2.50cm}>{\centering\arraybackslash} p{3cm}>{\centering\arraybackslash} p{3cm} >{\centering\arraybackslash} p{2.25cm} >{\centering\arraybackslash}p{1.25cm} 
>{\centering\arraybackslash}p{1.25cm} >{\centering\arraybackslash} p{2.25cm}}
\toprule
$F$ & $\omega_{r}/2\pi$ [GHz] & $\omega_p/2\pi$ [GHz] & $\detuning/ 2\pi $ [MHz] & $U/2\pi$ [kHz] & $L$ & $\kappa/2\pi$ [kHz]  & $G/2\pi$ [kHz]\\
\midrule
0.35 & 4.361407 & 8.722320 to 8.723800 & -0.5 to 0.25  & -5.58 & 1.64 & 70 & 302 \\

 0.52 & 4.349487  & 8.698474 to 8.699954 & -0.5 to 0.25  & -7.00 & 1.31 & 69  & 300 \\

 0.66 & 4.334886 & 8.669280 to 8.670760  &  -0.5 to 0.25  & -9.14 & 1 & 70 & 302 \\

 0.75 & 4.321900 & 8.643310 to 8.644790 & -0.5 to 0.25  & -11.36 & 0.80 & 75 & 301 \\

0.82 & 4.309052 & 8.617610 to 8.619090 & -0.5 to 0.25  & -13.86 &  0.66 & 76  & 293 \\
\bottomrule
\end{tabular}
\label{table:scaling_I}
\end{table*}

For the Scaling (II), we want to apply the following transformation:
\begin{align}
\label{Eq:Scaling2_supp}
     & (\text{II}): &  \!\!\!  \detuning&= \tilde{\detuning} L  &  G&=   \tilde{G} L  &   U&= \tilde{U}  &   \kappa&= \tilde{\kappa}L. 
\end{align}
In this case, the Kerr nonlinearity $U$ remains fixed, while the detuning $\detuning$ and the two-photon pump $G$ scale with the value of $L$. The Scaling (II) is implemented by fixing the external flux $F$ to 0.66, ensuring that $U$ and $\omega_r$ remain constant. Then, the two-photon pump $G$ is progressively increased, with its value determined by extrapolating $\average{\hat{n}_{\detuning, \rm ss}}$ to the x-intercept. The scaling of $G$ fixes the value of $L$. Finally, knowing $L$, the pump frequency $\omega_p$ is adjusted to maintain the scaling relation. Table \ref{table:scaling_II} summarizes the different operating points and the parameters used to realize the Scaling (II).

\begin{table*}
\centering
\caption{\textbf{Operating points and parameters used for Scaling (II)}}
\begin{tabular}{>{\centering\arraybackslash} p{1.25cm}
>{\centering\arraybackslash} p{2.50cm}>{\centering\arraybackslash} p{3cm}>{\centering\arraybackslash} p{3cm} >{\centering\arraybackslash} p{2.25cm} >{\centering\arraybackslash}p{1.25cm} 
>{\centering\arraybackslash}p{1.25cm} >{\centering\arraybackslash} p{2.25cm}}
\toprule
$F$ & $\omega_{r}/2\pi$ [GHz] & $\omega_p/2\pi$ [GHz] & $\detuning/2\pi$ [MHz] & $U/2\pi$ [kHz] & $L$ & $\kappa/2\pi$ [kHz]  & $G/2\pi$ [kHz]\\
\midrule
0.66 & 4.334898  & 8.670002 to 8.670557 & -0.1 to -0.38  & -9.14  & 1  & 70 & 291  \\
0.66 & 4.334893 & 8.670084 to 8.670881 &  -0.15 to -0.55 & -9.14 & 1.44 & 70 & 428 \\
0.66 & 4.334891 & 8.670193 to  8.671309 &  -0.21 to -0.76 & -9.14  & 1.99  & 70 & 594 \\
0.66 & 4.334896 & 8.670316 to 8.671791 & -0.26 to -1.00  & -9.14 & 2.66 & 70 &  796 \\
0.66 & 4.334895 & 8.670421 to 8.672202 &   -0.32 to -1.21 & -9.14 &  3.22 & 70 &  965 \\
0.66 & 4.334896 & 8.670523 to 8.672600 &  -0.37 to -1.40 & -9.14 & 3.77 & 70 &  1130 \\
0.66 & 4.334896 & 8.670617 to 8.673000 & -0.31 to -1.60  & -9.14  &  4.26 & 70 &  1278 \\
\bottomrule
\end{tabular}
\label{table:scaling_II}
\end{table*}

\begin{figure}
    \centering
    \includegraphics[width=\columnwidth]{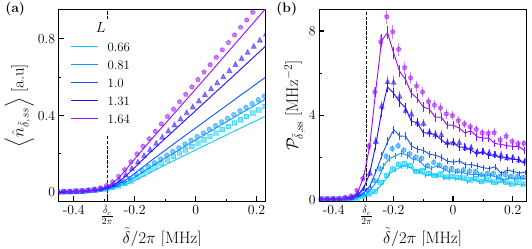}
    \caption{Comparison between the Scalings I in Eq.~\eqref{Eq:Scaling1} and II in Eq.~\eqref{Eq:Scaling2} (solid lines) : (a) Output photon number at the steady state
    vs scaled detuning $\tilde{\detuning}$ for increasing $L$. (b) Precision  $\mathcal{P}_{\tilde{\detuning}, {\rm ss} }$ as a function of $\tilde{\detuning}$.
    Parameters as in Fig.~\ref{fig:second_order_DPT}.
    }
    \label{fig:scaling_proof}
\end{figure}

The equivalence of the two scaling is demonstrated in Fig.~\ref{fig:scaling_proof}. In Fig.~\ref{fig:scaling_proof} (a), we compare $\average{\hat{n}_{\delta, {\rm ss}}}$ vs $\delta$ for scaling (I) and $\langle{\hat{n}_{\tilde{\delta}, {\rm ss}}} \rangle$ vs $\tilde{\delta}$ for scaling (II). 
Similarly in Fig.~\ref{fig:scaling_proof} (b), $\mathcal{P}_{\tilde{\detuning}, {\rm ss}}= \mathcal{P}_{{\detuning}, {\rm ss}}$ for Scaling (I) and $\mathcal{P}_{\tilde{\detuning}, {\rm ss}} = \mathcal{P}_{{\detuning}/L, {\rm ss}}$ for Scaling (II). We see that both scalings lead to similar results for both photon number and the precision.

\subsection{Precision estimation from the measured data}
\label{appendix_section:precision_estimation}

 After determining the parameters ($G$, $F$, and $\omega$), we proceed with the measurements.  For each measurement, the pump $G$ is turned on at a fixed frequency $\omega_p$ and we measure a trajectory by continously acquiring the signal quadratures $\hat I$ and $\hat Q$ for a total time time $T=\SI{69}{\micro\second}$. Each quadrature is integrated for a time $\Delta T$ of $\SI{1.5}{\micro\second}$ for all the steady state measurements (Figs. ~\ref{fig:second_order_DPT}, ~\ref{fig:scaling_I_enhancement}, ~\ref{fig:scalingII} and ~\ref{fig:scaling_proof}) and $\SI{250}{\nano\second}$ for the dynamical measurement [inset of Fig.~\ref{fig:second_order_DPT}(b)]. Mathematically, we define 

\begin{equation}
\begin{split}
\hat{I}_j = \frac{1}{\Delta T} & \int_{t_j}^{t_j+\Delta T} d\tau  \sqrt{\mathcal{G}} \left[ \hat{i}(\tau) +  \hat{\mu}(\tau)   \right], \\ &(t_{j+1}= t_j +\Delta T) \quad (j=1,...,J)
\end{split}
\end{equation}

\begin{equation}
\begin{split}
\hat{Q}_j = \frac{1}{\Delta T} &\int_{t_j}^{t_j+\Delta T} d\tau \sqrt{\mathcal{G}} \left[   \hat{q}(\tau) +   \hat{\nu}(\tau)   \right], \\ &(t_{j+1}= t_j +\Delta T) \quad (j=1,...,J) 
\end{split}
\end{equation}

where $\mathcal{G}$ represents the effective gain of the output line, $ \hat{\mu}(\tau)$ and $\hat{\nu}(\tau)$ are the amplifier noise quadratures, and $  \hat{i}(\tau)$ and $\hat{q}(\tau)$ are the signal quadrature. Notice that $\hat{I}_j$ and $\hat{Q}_j$ are incompatible observables. However, they can be approximately measured simultaneously in the large $\mathcal{G}$ limit. The index $j$ ranges from 1 to $J=T/\Delta T$ and corresponds to the number of data points acquired in a measurement trace.  To simplify the notation, we define the following
\begin{align}
      \hat{i}_j &= \frac{1}{\Delta T} \int_{t_j}^{t_j+\Delta T} d\tau\,   \hat{i}(\tau)\\
      \hat{q}_j &= \frac{1}{\Delta T} \int_{t_j}^{t_j+\Delta T} d\tau\,   \hat{q}(\tau)\\
      \hat{\mu}_j &= \frac{1}{\Delta T} \int_{t_j}^{t_j+\Delta T} d\tau\,   \hat{\mu} (\tau)\\ 
      \hat{\nu}_j &= \frac{1}{\Delta T} \int_{t_j}^{t_j+\Delta T} d\tau\, \hat{\nu} (\tau).
\end{align}

Using these definitions, the quantity of interest - i.e., the measured output power - is 
\begin{equation}
      \hat{N}_j =  \hat{I}_j^2+ \hat{Q}_j^2 = \mathcal{G} \left[   \hat{i}_j^2+  \hat{q}_j^2+ \hat{\mu}_j^2+ \hat{\nu}_j^2 + 2  \hat{\mu}_j  \hat{i}_j+2  \hat{\nu}_j  \hat{q}_j\right ].
\end{equation}

After the single measurement trace has been collected, the pump $G$ is turned off and we wait $\SI{100}{\micro \second}$ for the resonator to return to the vacuum state before acquiring the next trace. The precision for estimating the parameter $\detuning$ depends on the first and second moments of $\langle\hat{N}_{j}\rangle$ at two different, but close, values of $\detuning$. 
The moments are estimated by repeating the measurement $M=400\times 10^3$ times to obtain the set $\{I_{j,m},Q_{j,m}\}_{m=1}^M$ of measurement samples. This allows us to estimate the expectation values
\begin{equation}
\label{N:first_moment}
    \average{\hat N_j} \approx  \frac{1}{M} \sum_{m=1}^M   I_{j,m}^2+  Q_{j,m}^2
\end{equation}
\begin{equation}
\label{N:second_moment}
    \average{\hat N^2_j} \approx  \frac{1}{M} \sum_{m=1}^M (  I_{j,m}^2+  Q_{j,m}^2)^2
\end{equation}

The measurement protocol is repeated for different values of detuning $\detuning$ to obtain $\average{\hat N_{\delta,j}}$ and $\average{\hat N^2_{\delta,j}}$. We can then calculate the error on the estimation of $\delta$ using measurements with nearby detuning values (i.e, separated by a small $\epsilon)$
\begin{equation}
\label{eq_supp:estimation}
        \error{\detuning}_{\rm ss}  = \frac{\standarddeviation{N_{\detuning,{\rm ss}}}}{\left|\partial_{\detuning} 
    \average{\hat{N}_{\detuning, {\rm ss}}} \right|} \simeq 
    \frac{\left(\standarddeviation{N_{\detuning,{\rm ss}}} + \standarddeviation{N_{\detuning+\discretization,{\rm ss}}}\right) \discretization}{2 \left|\average{\hat{N}_{\detuning, {\rm ss}}}- \average{\hat{N}_{\detuning + \discretization, {\rm ss}}}\right| }
\end{equation}
where $\Delta N_{\delta,j}=\sqrt{ \average{\hat N^2_{\delta,j}} -\average{\hat N_{\delta,j}}^2} $. Finally, the precision of the estimation of $\detuning$ is simply  $\mathcal{P}_{\detuning, j} =    \left( \error{\detuning _j }\right)^{-2}$. The steady state precision is calculated with greater accuracy by averaging over the $J_{{\rm ss}}$ values of precisions where the system has reached the steady state. We label the index $j_{{\rm ss}}$ as the first index where the system is in the steady state ($J_{\rm ss}= J-j_{\rm ss})$
\begin{equation}
    \mathcal{P}_{\detuning, {\rm ss}} = \frac{1}{J_{\rm ss}} \sum_{j=j_{\rm ss}}^J \mathcal{P}_{\detuning, j}
\end{equation}
We find that the system has well reached the steady state for $t>\SI{15}{\micro\second}$, which corresponds to approximately seven times the dissipation rate. 

\begin{figure}[h]
        \centering
\includegraphics[width=\columnwidth]{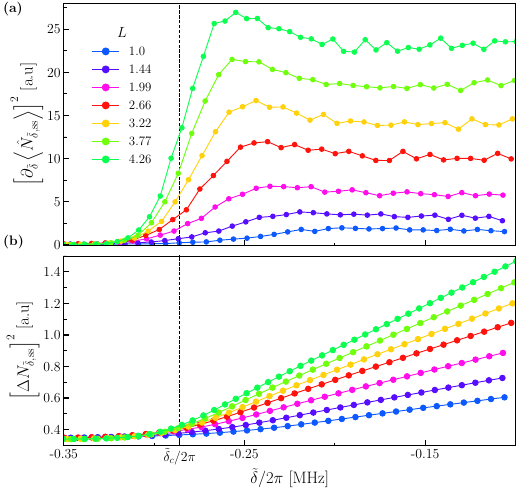}
        \caption{For the Scaling II, (a) signal  $\left[\partial_{\tilde{\detuning}} 
    \average{\hat{N}_{\tilde{\detuning}, {\rm ss}}} \right]^2$ and (b) noise $\left[ \Delta N_{\tilde{\delta},\rm{ss}} \right]^2$ vs scaled detuning $\tilde{\delta}$ for increasing $L$.  } 
        \label{Fig:terms_precision}
\end{figure}

For clarity, we separately show in Fig. \ref{Fig:terms_precision}, the numerator and denominator of Eq. \ref{eq_supp:estimation}, which are responsible for the enhancement in precision near the critical point. The ratio of these two quantities corresponds to the measured precision, $\mathcal{P}_{\delta,\rm{ss}}$.

Finally, notice that we can parameterized the expectation value of the output power as $\langle \hat N_j\rangle=\mathcal{G}[\langle  \hat n_j\rangle+n_{\rm amp}]$, as done in the main text. We can obtain an estimation of $\mathcal{G}n_{{\rm{amp}}} $ by performing the same type of measurement, but with $G=0$, such that the intracavity state is vacuum.  This allows us to obtain $\expval{\hat n_j}$ up to the proportionality factor $\mathcal{G}$ for Figs.~\ref{fig:second_order_DPT}(a) and \ref{fig:scalingII}(a). 
\\
\subsection{Precision dependence on the amplifier noise}
Let us estimate the precision defined Eq.~\eqref{eq_supp:estimation} on the amplifier noise. We model the amplifier noise as white thermal noise with $\langle \hat \mu (\tau) \hat \mu (\tau')\rangle=\langle \hat \nu (\tau)\hat \nu (\tau')\rangle=\sigma^2\delta(t-t')$ and $\langle \hat \mu (\tau)\hat\nu (\tau')\rangle=0$, where $\sigma^2\geq 1/4$ (we have set the variance of the vacuum quadrature to $1/4$). This implies that $\langle \hat \mu_j^2\rangle = \langle \hat \nu_j^2\rangle = \sigma^2/\sqrt{\Delta T}$ and $\langle \hat \mu_j\hat \nu_j\rangle=0$, where $\Delta T$ is the integration time. 

Let us also define the power operators $\hat p_j = \sqrt{\Delta T}(\hat i_j^2+\hat q_j^2)$. 
We have that 
\begin{align}
\frac{1}{\mathcal{G}}\langle \hat N_j\rangle = \langle \hat i_j^2\rangle+    \langle \hat q_j^2\rangle+  \langle \hat \mu_j^2\rangle+\langle\hat \nu_j^2\rangle = \frac{1}{\sqrt{\Delta T}}\left[\langle \hat p_j\rangle+2\sigma^2\right].
\end{align}
For the second moment, we get

\begin{equation}
\begin{split}
\frac{1}{\mathcal{G}^2} \langle \hat N_j^2\rangle =  &\frac{1}{\Delta T}\langle \hat p_j ^2\rangle + \langle (\hat \mu_j^2+\hat \nu_j^2)^2\rangle + 4\langle \hat \mu_j^2\rangle\langle \hat i_j^2\rangle \\&+4\langle \hat \nu_j^2\rangle\langle  \hat q_j^2\rangle+2\langle \hat \mu_j^2+\hat \nu_j^2\rangle\langle \hat i_j^2+\hat q_j^2\rangle
\end{split}
\end{equation}
\begin{align}
\quad& = \frac{1}{\Delta T} \langle \hat p_j ^2\rangle + \langle (\hat \mu_j^2+\hat \nu_j^2)^2\rangle + \frac{8\sigma^2 }{\Delta T}\langle \hat p_j\rangle \\
\quad& = \frac{1}{\Delta T} \left[ \langle \hat p_j ^2\rangle +8\sigma^4 -1/4+ 8\sigma^2 \langle \hat p_j\rangle\right],
\end{align}

where we have used that $\langle(\hat \mu_j^2+\hat \nu_j^2)^2\rangle = (8\sigma^4-1/4)/\Delta T$. All in all, we obtain
\begin{equation}
\begin{split}
\frac{\Delta N_j}{|\partial_\delta \langle \hat N_j\rangle|} =&\frac{\Delta p_j+4\sigma^2\langle \hat p_j\rangle +4\sigma^4-1/4}{|\partial_\delta \langle \hat p_j\rangle|} \\=& \frac{\Delta p_j}{|\partial_\delta \langle \hat p_j\rangle|}\left[1+\frac{4\sigma^2\langle \hat p_j\rangle+4\sigma^4-1/4}{\Delta p_j}\right],
\end{split}
\end{equation}
where $\Delta p_j = \sqrt{\langle \hat p_j^2\rangle-\langle \hat p_j\rangle^2}$. Since  $\langle \hat p_j\rangle > 0$ and $4\sigma^4-1/4\geq0$, it is clear that increasing the amplifier noise level $\sigma^2$ induces a decrease of precision in the metrological protocol.

\subsection{Error bars}
The error on the estimation of the precision, which we label ${\rm Err}[P_{\delta,j}]$ to simplify the notation, is calculated using the error propagation formula applied to Eq.~\eqref{eq_supp:estimation}
\begin{equation}
\begin{split}
&{\rm Err}[P_{\delta,j}]=  2 P_{\delta,j} \left( \frac{{\rm Err}\left[\expval{\hat{N}_{\delta,j}} \right]^2+  {\rm Err}\left[\expval{\hat{N}_{\delta+\epsilon,j}}\right] ^2}{\left( \expval{\hat{N}_{\delta,j}}-\expval{\hat{N}_{\delta+\epsilon,j}} \right)^2}\right. \\ & \left.+ \frac{{\rm Err}[\Delta N_{\delta,j}]^2+{\rm Err}[\Delta N_{\delta+\epsilon,j}]^2}{\left( \Delta N_{\delta,j}-\Delta N_{\delta+\epsilon,j} \right)^2}+\frac{{\rm Err}[\epsilon]^2}{\epsilon^2}  \right)^{1/2}
\end{split}
\end{equation}
%
where the error on the sample means are given by ${\rm Err}[\average{\hat{N}_{\delta,j}}]=\sqrt{\Delta N_{\delta,j}^2/M}$ and ${\rm Err}[\average{\hat{N}_{\delta+\delta,j}}]=\sqrt{\Delta N_{\delta+\delta,j}^2/M}$, while the errors on the standard deviations are given by ${\rm Err}[\Delta N_{\delta,j}]=\sqrt{\Delta N_{\delta,j}^2/(2M)}$ and ${\rm Err}[\Delta N_{\delta+\epsilon,j}]=\sqrt{\Delta N_{\delta+\epsilon,j}^2/(2M)}$. The error in the detuning step $\epsilon$ comes from fluctuations in the resonator frequency $\omega_r$. By continuously monitoring the resonator frequency through spectroscopy measurements, we estimate that the resonator frequency fluctuations over a time duration comparable to the measurement time are at most on the order of \SI{1}{\kilo\hertz}. The final error for the steady state value is then taken as 
\begin{equation}
\begin{split}
    {\rm Err}[P_{\delta,{\rm ss}}] &=  \sqrt{\frac{ \left( \frac{1}{J_{\rm ss}} \sum_{j=j_{\rm ss}}^J {\rm Err}[P_{\delta,j}] \right)^2}{J_{ss}}} \\ &\approx \sqrt{ \frac{{\rm Err}[P_{\delta,j_{ss}}]^2}{J_{ss}}}
\end{split}
\end{equation}

\section{Classical benchmark}
\label{appendix:classical}

Here, we provide the details about the parameter estimation protocol based on a linear driving. We will demonstrate two facts: (i) The maximal precision for a linear resonator is achieved when a pumping mechanism drives it at resonance; (ii) The precision increases linearly with the number of photons. Using the language of the main text, we deduce that $P_{\delta_{\rm max}, \rm ss} \propto \langle \hat{a}^\dagger \hat{a} \rangle$, the photon number in the resonator, with $\delta_{\rm max}=0$.

\subsection{Estimation of the precision with single frequency-resolved measurement}
We consider a one-sided resonator driven with a monochromatic coherent input centered around the frequency $\omega_p$. In this case, information about the resonator frequency is imparted on the phase of the reflected signal, which is read out throught homodyne measurement. 
Let us define the input field as
\begin{equation}
|\psi_{in}\rangle = \exp \left\{ \alpha^*\ \hat{m} - \alpha\ \hat{m}^\dagger \right\} |0\rangle \equiv \hat{D}_m(\alpha) |0\rangle,
\end{equation}
where $\hat{D}_m(\alpha)$ is a displacement operator of the single mode $\IN{m}$, defined as
\begin{equation}
\IN{m} = \int d\omega\ f(\omega)\ \IN{a}(\omega),\qquad \text{where} \qquad
\int d\omega f(\omega)^2=1.
\end{equation}
For the sake of simplicity, we take $f(\omega)\in \mathbb{R}$.
Using input-output theory we find a linear relationship between the frequency components of the input and output fields,
\begin{equation}
\OUT{a}(\omega) = 
\frac{\kappa_{\rm ext}/2 + i(\omega - \omega_r) }{\kappa_{\rm ext}/2 - i(\omega - \omega_r) } \IN{a}(\omega)
\equiv \Gamma(\omega) \IN{a}(\omega).
\end{equation}
We assume for now that we are able to measure an output mode which has the same profile as the input wavepacket,
\begin{equation}
\OUT{m} \equiv \int d\omega\ f(\omega) \OUT{a}(\omega) = \int d\omega\ f(\omega) \Gamma(\omega) \IN{a}(\omega).
\end{equation}
Given that the input is a Gaussian state, we know that the optimal measurement is homodyne. In order to estimate the precision over the estimation of $\omega_r$, we need to calculate the first- and second-moments of the readout mode $\OUT{m}$.

Using the relation $\IN{a}(\omega) \hat{D}_m(\alpha)|0\rangle= \alpha f(\omega) \hat{D}_m(\alpha)|0\rangle$, we find for the first moment of the output mode
\begin{equation}
\langle\OUT{m}\rangle= \alpha \int d\omega f(\omega)^2 \Gamma(\omega) \approx \alpha \Gamma(\omega_p),
\end{equation}
where we have taken the input to be effectively in a monochromatic wavepacket centered around the drive frequency $\omega_p$, and with a bandwidth much smaller than the resonator linewidth $\kappa_{\rm ext}$ and the drive frequency $\omega_p$, that is
\begin{equation}
\begin{split}
&\int d\omega\ \omega f(\omega)^2 = \omega_p,\\ &\text{and} \quad
\sqrt{\int d\omega \left(\omega^2 - \omega_p^2\right) f(\omega)^2} \equiv \Delta \omega \ll \kappa_{\rm ext},\omega_p.
\end{split}
\end{equation}
The condition $\Delta \omega\ll\omega_p$ is the narrow bandwidth approximation, which allows us to work with integration limits from $-\infty$ to $+\infty$.
Similarly, we can calculate the second-order moment of the output mode, as for example:
\begin{align} 
&\langle \OUT{m}^\dagger \OUT{m}\rangle = 
\nonumber \\ \nonumber
&\iint d\omega_1 d\omega_2 \ f(\omega_1) f(\omega_2) \Gamma(\omega_1)^* \Gamma(\omega_2) 
\langle \IN{a}(\omega_1)^\dagger \IN{a}(\omega_2)\rangle
\\ \nonumber
& =|\alpha|^2 \iint d\omega_1 d\omega_2 \ f(\omega_1)^2 f(\omega_2)^2 \Gamma(\omega_1)^* \Gamma(\omega_2)
\\ 
& =|\alpha|^2 |\Gamma(\omega_p)|^2 = |\alpha|^2.
\end{align}

We want to estimate $\omega_r$ from homodyne measurements performed on the output mode $\OUT{m}$. Let us define $\omega_r = \omega_0 + \delta$, where $\omega_0$ is our prior information on the parameter and $\delta$ is a small shift to be estimated. To simplify the expressions, we define $\Delta_p = \omega_p - \omega_0$, as the detuning between the driving and the prior frequency. We then rewrite
\begin{eqnarray}
\Gamma(\omega_p) &=&
\frac{\kappa_{\rm ext}/2 + i(\omega_p - \omega_r) }{\kappa_{\rm ext}/2 - i(\omega_p - \omega_r) } 
\nonumber \\ 
&=& \frac{\kappa_{\rm ext}/2 + i(\Delta_p - \delta) }{\kappa_{\rm ext}/2 - i(\Delta_p - \delta) }\equiv \Gamma(\Delta_p).
\end{eqnarray}
Notice that since $|\Gamma(\Delta_p)|=1$, then we can rewrite
\begin{equation}
\begin{split}
\Gamma(\Delta_p) =&
\left(
\frac{\kappa_{\rm ext}/2 + i \Delta_p }{\kappa_{\rm ext}/2 - i\Delta_p }
\right)\times \\
\times& \left(
\frac{\kappa_{\rm ext}^2 - 4\Delta_p \delta + 4\Delta_p^2 - 2i\kappa_{\rm ext}\delta}{\kappa_{\rm ext}^2 - 4\Delta_p \delta + 4\Delta_p^2 + 2i\kappa_{\rm ext}\delta}
\right) \\
=&
e^{i\phi_0}\ \left[
1-\frac{4i\kappa_{\rm ext} \delta}{\kappa_{\rm ext}^2 + 4\Delta_p^2} + O(\delta^2)
\right],
\end{split}
\end{equation}
where $e^{i \phi_0}$ is a phase, which is independent of the parameter $\delta$ to be estimated. We have expanded the expression up to the first order in $\delta$ to make explicit the value of the first-order derivative. This expression helps us identify the phase of the optimal readout quadrature, given by
$\OUT{p} = \left(e^{-i \phi_0} \OUT{m} -  e^{i \phi_0} \OUT{m}^\dagger \right)/2i$.

The precision over the estimation of $\omega_r$ is then given by the precision of the homodyne signal,
\begin{equation}
\label{SNR_hom}
\mathcal{P}(\Delta_p) = \frac{\left[\partial_{\delta} \langle \OUT{p}(\phi)\rangle\right]^2}{\langle \Delta \OUT{p}(\phi)^2\rangle}.
\end{equation}
As the mode $\OUT{m}$ is in a coherent state, we have directly $\langle\Delta\OUT{p}(\phi)^2\rangle\equiv   \langle\OUT{p}(\phi)^2\rangle- \langle \OUT{p}(\phi)\rangle^2 =1/4$.
For the signal we find
\begin{equation}
\begin{split}
\langle \OUT{p}(\phi)\rangle = & \frac{\alpha}{2i}\left[\Gamma(\Delta_p)e^{-i\phi_0} - \Gamma(\Delta_p)^*e^{i\phi_0}\right] \\ =& \alpha \Im\left\{\Gamma(\omega_p) e^{-i\phi_0}\right\}\\ =&-  \frac{4\kappa_{\rm ext} }{\kappa_{\rm ext}^2 + 4\Delta_p^2} \alpha \delta + O(\delta^2).
\end{split}
\end{equation}
Inserting this result into Eq.~\eqref{SNR_hom}, we finally find
\begin{equation}
\label{precision_no_t}
\mathcal{P}(\Delta_p) = \frac{64 \kappa_{\rm ext}^2}{\left(\kappa_{\rm ext}^2 + 4\Delta_p^2\right)^2}\alpha^2.
\end{equation}
The precision reaches its maximum for $\Delta_p=0$, and is linear in the number of photons of the output field $\langle \OUT{m}^\dagger \OUT{m}\rangle=\alpha^2$.

\subsection{Precision dependence on measurement time and bandwith}
So far, we have assumed that we can measure the output mode $\OUT{m}$. This mode is narrowly peaked in frequency and, accordingly, it has a wide distribution in time. Hence, in deriving Eq.~\eqref{precision_no_t} we have implicitly assumed a long measurement time. 
In order to calculate the time scaling of the estimation precision, we must define readout modes accessible with a finite measurement time $T$, 
\begin{align}
\hat A_{\rm out} = \frac{1}{\sqrt{T}}\int_{t}^{t+T}d\tau\ e^{i\omega_p(\tau-t_0)}\hat a_{\rm out} (\tau),
\end{align}
where $\hat a_{\rm out} (\tau)= \frac{1}{\sqrt{2\pi}}\int d\omega \ e^{-i\omega(\tau-t_0)}\hat a_{\rm out}(\omega)$.
Let us calculate the first and second moments of this mode. As frequency distribution of the drive, we consider a narrow square function $f= \frac{1}{\sqrt{B}}\mathbb{I}[\omega\in(\omega_p-B/2,\omega_p+B/2)]$. We have that
\begin{eqnarray}
&\langle& \hat A_{\rm out}\rangle= \frac{1}{\sqrt{T}}\int_t^{t+T} d\tau\ e^{i\omega_p(\tau-t_0)}\langle\hat a_{\rm out} (\tau)\rangle 
\nonumber \\ 
&=& \frac{1}{\sqrt{2\pi T}}\int_t^{t+T} d\tau\ e^{i\omega_p(\tau-t_0)}\int d\omega \ e^{-i\omega(\tau-t_0)}\langle\hat a_{\rm out}(\omega)\rangle
\nonumber \\
 &=& \frac{1}{\sqrt{2\pi T}}\int_t^{t+T} d\tau\ e^{i\omega_p(\tau-t_0)}\int d\omega \ e^{-i\omega(\tau-t_0)}\Gamma(\omega)f(\omega)\alpha 
 \nonumber \\
&\approx & \frac{1}{\sqrt{2\pi T}}\int_t^{t+T} d\tau\ e^{i\omega_p(\tau-t_0)}e^{-i\omega_p(\tau-t_0)} \times
\nonumber \\
&\times& \int_{\omega_p-B/2}^{\omega_p+B/2} d\omega\ \frac{\Gamma(\omega_p)\alpha }{\sqrt{B}} =
\nonumber \\
&=& \sqrt{\frac{BT}{2\pi}}\Gamma(\omega_p)\alpha.
\end{eqnarray}
Similarly, we get that $N_{\rm out} = \langle \hat A_{\rm out}^\dag \hat A_{\rm out}\rangle = BT |\alpha|^2/2\pi$. Since $|\langle \hat A_{\rm out}\rangle|=\sqrt{\langle\hat A_{\rm out}^\dag \hat A_{\rm out}\rangle}$, where we use that $|\Gamma(\omega_p)|=1$, we conclude that $\hat A_{\rm out}$ is in a coherent state, and the variance of its quadratures is $1/4$. The precision of the optimal observable, which is $\hat P_{\rm out} =( e^{-i\phi_0}\hat A_{\rm out} -e^{-i\phi_0}\hat A_{\rm out}^\dag ) /2 i$, is
\begin{align}
\mathcal{P}_{\Delta_p,T} =\frac{64 \kappa_{\rm ext}^2}{\left(\kappa_{\rm ext}^2 + 4\Delta_p^2\right)^2} \frac{BT\alpha^2}{2\pi}
= \frac{64 \kappa_{\rm ext}^2}{\left(\kappa_{\rm ext}^2 + 4\Delta_p^2\right)^2}N_{\rm out},
\end{align}
which is optimal for $\Delta_p=0$.
To conclude, when the input is a monochromatic coherent state and the resonator has no quantum nonlinearity, the optimal scaling of the estimation precision is linear with respect to time and photon number.

\end{document}